\documentclass[aps,prb,reprint,nobibnotes,amsmath,amssymb,showpacs,floatfix,superscriptaddress,bibliography,nofootinbib]{revtex4-1}
\usepackage{amsmath,empheq}
\usepackage{amsthm}
\usepackage{amsfonts}
\usepackage{amssymb}
\usepackage{amsxtra}
\usepackage{mathtools}
\usepackage{xcolor}
\usepackage{graphicx}
\usepackage{subfigure}
\usepackage{dcolumn}
\usepackage{mathrsfs}
\usepackage{graphicx,wrapfig,lipsum}
\usepackage{tikz}
\usepackage{float}
\usepackage{bm}
\usepackage{siunitx}
\usepackage[breaklinks=true,colorlinks,citecolor=blue,linkcolor=blue,urlcolor=blue]{hyperref}

\usepackage{comment}

\DeclareMathAlphabet{\bi}{OML}{cmm}{b}{it}
\def\be{\begin{equation}}
	\def\ee{\end{equation}}
\def\bearr{\begin{eqnarray}}
	\def\eearr{\end{eqnarray}}

\begin{document}
	\title{Terahertz anomalous Hall effect in magnetic Weyl semimetal Co$_3$Sn$_2$S$_2$} 
	\author{Ashutosh Singh}
	\email{asingh.n19@gmail.com}
    \thanks{Current address:
        School of Physics and Optoelectronic Engineering, Hainan University, Haikou 570228, China}
	\affiliation{Department of Physics and Astronomy, Texas A\&M University, College Station, TX 77843, USA}

\author{Hongjing Xu}
\affiliation{Department of Physics and Astronomy, Rice University, Houston, TX 77005, USA}
\author{Xielin Wang}
\affiliation{Applied Physics Graduate Program, Smalley-Curl Institute, Rice University, Houston, TX 77005, USA}                
\affiliation{Department of Electrical and Computer Engineering, Rice University, Houston, TX 77005, USA}
\author{Kohei Fujiwara}
\affiliation{Institute for Materials Research, Tohoku University, Sendai 980-8577, Japan}
\affiliation{Department of Chemistry, Rikkyo University, Tokyo 171-8501, Japan}
\author{Atsushi Tsukazaki}
\affiliation{Institute for Materials Research, Tohoku University, Sendai 980-8577, Japan}
\affiliation{Department of Applied Physics and Quantum-Phase Electronics Center (QPEC),
The University of Tokyo, Tokyo 113-8656, Japan}
\author{Shengxi Huang}
\affiliation{Department of Electrical and Computer Engineering, Rice University, Houston, TX 77005, USA}
\affiliation{Department of Materials Science and NanoEngineering, Rice University, Houston, TX 77005, USA}
\affiliation{Smalley-Curl Institute, Rice University, Houston, TX 77005, USA}
\author{Andrey Baydin}
\affiliation{Department of Electrical and Computer Engineering, Rice University, Houston, TX 77005, USA}
\affiliation{Smalley-Curl Institute, Rice University, Houston, TX 77005, USA}
\author{Junichiro Kono}
\affiliation{Department of Physics and Astronomy, Rice University, Houston, TX 77005, USA}
\affiliation{Department of Electrical and Computer Engineering, Rice University, Houston, TX 77005, USA}
\affiliation{Department of Materials Science and NanoEngineering, Rice University, Houston, TX 77005, USA}
\affiliation{Smalley-Curl Institute, Rice University, Houston, TX 77005, USA}
\author{Alexey Belyanin}%
\email{belyanin@tamu.edu}
\affiliation{Department of Physics and Astronomy, Texas A\&M University, College Station, TX 77843, USA}

\begin{abstract}
    Time-reversal-symmetry-broken Weyl semimetals are known to have at least two nodes in their electronic band structure, separated in momentum space and acting as sources and sinks of Berry curvature. This gives rise to a transverse Hall conductivity, known as the anomalous Hall effect (AHE), which, in the simplest two-node picture, is proportional to the momentum-space separation between the nodes in the zero frequency limit. In the recently discovered Weyl semimetal $\mathrm{Co_3Sn_2S_2}$, a giant AHE has been observed. However, experimental investigations in the low-energy regime, which directly probe quasiparticle excitations near the Weyl nodes, remain limited. Here, we present a systematic study of the intrinsic low-energy gyrotropic optical response of $\mathrm{Co_3Sn_2S_2}$ using terahertz spectroscopy combined with semianalytical calculations based on a physically intuitive effective model. Our results provide a robust and transparent explanation of the observed magnetooptical phenomena in terms of intrinsic gyrotropy arising from momentum-space separation of the Weyl nodes. Furthermore, quantitative comparison between experiment and theory places stringent constraints on the material parameters.
\end{abstract}
	
	%
	\maketitle
	%
\section{Introduction}
Much of the recent progress in the study of topological phenomena in condensed-matter systems has been driven by the realization of Weyl fermions in three-dimensional solids.\cite{Ashvin1,Ashvin2,Burkov} Breaking either time-reversal or inversion symmetry in a three-dimensional Dirac semimetal gives rise to pairs of Weyl nodes, around which the quasiparticle dispersion is linear.\cite{Zyuzin_2012} It has been confirmed experimentally that Weyl fermions exist in several materials, including TaAs, NbAs, and NbP.\cite{TaAs, TaAs1,lv2015experimental, NbAs, shekhar2015extremely} Among the most remarkable optical properties of Weyl semimetals are their intrinsic magneto-optical responses, which originate from a finite transverse conductivity in the absence of an external magnetic field and manifest themselves through the rotation of the polarization plane of reflected or transmitted light. This intrinsic response is associated with the divergence of the Berry curvature near Weyl points of opposite chirality. If the two nodes are separated in momentum space by a distance ${\bf Q}$, it has been shown that the zero-frequency limit of the real part of the Hall conductivity is given by $e^2|{\bf Q}|/(\pi h)$.\cite{Burkov} The leading finite-frequency correction to this constant term has been shown to scale quadratically with frequency.\cite{Trivedi,zyuzin}

Subsequent theoretical studies of the broadband optical properties of magnetic Weyl semimetals showed that the vector connecting two Weyl nodes in momentum space defines the axis of gyrotropy in real space, giving rise to gyrotropic and biaxially anisotropic bulk and surface optical conductivity tensors.\cite{PhysRevB.99.075137} Consequently, a variety of strong magneto-optical effects can occur in the absence of an external magnetic field, for both bulk and surface polariton modes. These include the intrinsic Faraday effect, the optical Hall effect, and various magnetoplasmon resonances arising from both intraband and interband transitions.\cite{PhysRevB.99.075137,chen2019}  

\begin{figure}[t!]
\includegraphics[width = \linewidth]{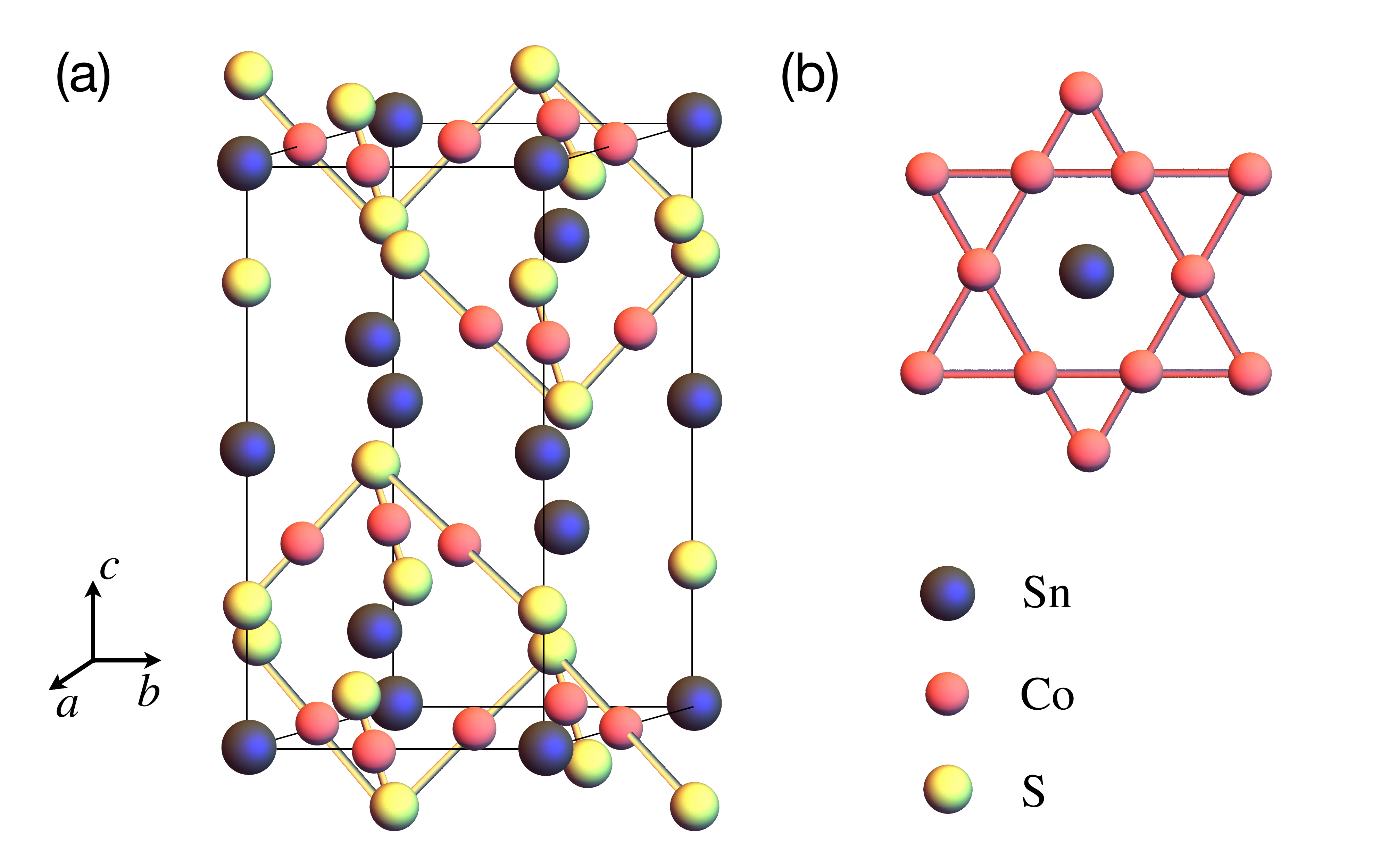}
\caption{(a) Layered structure of Co$_3$Sn$_2$S$_2$ and (b) two dimensional Co$_3$Sn kagome layer. Reproduced from Fig.~1(a) of Ref.~\cite{xing2020}. }
\label{fig0}
\end{figure}

Recently, a magnetic Weyl semimetal phase has been reported in the Kagome-lattice Co$_3$Sn$_2$S$_2$, which exhibits a remarkably high intrinsic Hall response and a high Curie temperature.\cite{Liu2018, Chen} In this material, the easy axis of magnetization lies along the $c$ axis owing to the out-of-plane magnetic moments of $\mathrm{Co}$ atoms in the kagome layers (Fig.~\ref{fig0}). Since its discovery, a variety of intriguing properties associated with its Weyl semimetal phase have been investigated, including Fermi arcs,\cite{Morali} one-dimensional chiral edge states,\cite{Howard} the zero-field Nernst effect,\cite{Guin} chirality switching in thin films in the terahertz (THz) range,\cite{Yoshikawa} and light-induced anomalous Hall conductivity.\cite{PhysRevB.111.245104} 
In Ref.~\onlinecite{Liu2018}, Co$_3$Sn$_2$S$_2$ was shown to exhibit a giant anomalous Hall effect (AHE). These strong magnetic and gyrotropic responses are believed to originate from spin-orbit coupling (SOC), which, in the presence of ferromagnetic order, breaks time-reversal symmetry (TRS) and splits a nodal ring into three pairs of Weyl nodes. \textit{Ab~initio} calculations further suggest that these nodes are located approximately $60$\,meV away from the chemical potential. A subsequent theoretical study of the intrinsic AHE in Co$_3$Sn$_2$S$_2$~\cite{Wang2018} reached similar conclusions and demonstrated that the Weyl nodes are robust, with a momentum-space separation of $0.3\,\mathrm{\mathring{A}}^{-1}$.  
Furthermore, broadband spectroscopic measurements spanning the THz/DC to the near-infrared range have revealed a giant magneto-optical response of topological origin.\cite{Okamura2020}

These studies highlight the importance of a systematic investigation of the optical response in the THz range, which provides valuable insight into the low-energy electronic excitations in the vicinity of the Weyl nodes as well as the underlying material parameters. 
In this paper, we present a combined theoretical and experimental study of the THz optical response and employ an effective low-energy model to provide an intuitive physical interpretation of the observed phenomena. Our model not only reproduces the experimentally observed behavior of the longitudinal conductivity and Faraday angle in the (sub-)THz frequency range, but also explains the sign reversal of the real part of the anomalous Hall conductivity and the decrease in its imaginary part above the frequency corresponding to twice the chemical potential.

The remainder of this paper is as follows. In Section~\ref{Dispersion}, we introduce the two-band model Hamiltonian and its eigenstates and eigenenergies. In Section~\ref{Op_con}, we calculate the optical conductivity and compare the real part of the longitudinal conductivity with experimental data. In Section~\ref{Farday_angle}, we compare the calculated and measured real part of the Faraday angle at different temperatures. Finally, in Section~\ref{Conclusion} we summarize our main findings. 

\section{Dispersion}\label{Dispersion}
A minimal model for the TRS breaking Weyl semimetal in the low-energy picture is given by the Hamiltonian~\cite{PhysRevB.89.235315}  
\begin{align}\label{ham}
\hat{\mathcal{H}} = t_1a(k_x\sigma_x + k_y\sigma_y) +  \left[t_2a^2(k_x^2+k_y^2+k_z^2)-b\right]\sigma_z,
\end{align}
where $t_1, t_2$, and $b$ are energy parameters explained below, $a$ is the lattice parameter so that $at_1/\hbar$ and $at_2/\hbar$ have the dimension of Fermi velocity \cite{RevModPhys.81.109}, and $\sigma_x$, $\sigma_y$, and $\sigma_z$ are Pauli matrices. In the cylindrical polar representation, taking $k_x\to k_p\cos\phi_{\bf k}$ and $k_y\to k_p\sin\phi_{\bf k}$, where $k_p=\sqrt{k_x^2+k_y^2}$ and $\phi_{\bf k}=\tan^{-1}(k_y/k_x)$, we have 
\begin{align}
\hat{\mathcal{H}} = \left(
\begin{array}{cc}
 t_2a^2(k_p^2+k_z^2)-b & t_1 a k_p e^{-i \phi_{\bf k} } \\
 t_1 ak_p e^{i \phi_{\bf k} } & b-t_2a^2(k_p^2+k_z^2) \\
\end{array}
\right)~.
\end{align}
In this picture, the energies of the conduction ($\lambda =+1$) and valence bands ($\lambda =-1$) are given by
\begin{align}
\varepsilon_{\lambda\bf k} = \lambda\sqrt{\left[t_2a^2(k_p^2+k_z^2)-b\right]^2 + t_1^2a^2k_p^2}~.
\end{align}
The optical transition energies are $\hbar\omega_{\bf k} = \varepsilon_{+\bf k}-\varepsilon_{-\bf k}$. The location of the Weyl points is obtained by setting $\varepsilon_{\lambda\bf k}=0$, such that the coordinates of the points in the Brillouin zone are $\left(0,0,\pm a^{-1}\sqrt{b/t_2}\right)$.
\begin{figure}[t!]
\includegraphics[width = \linewidth]{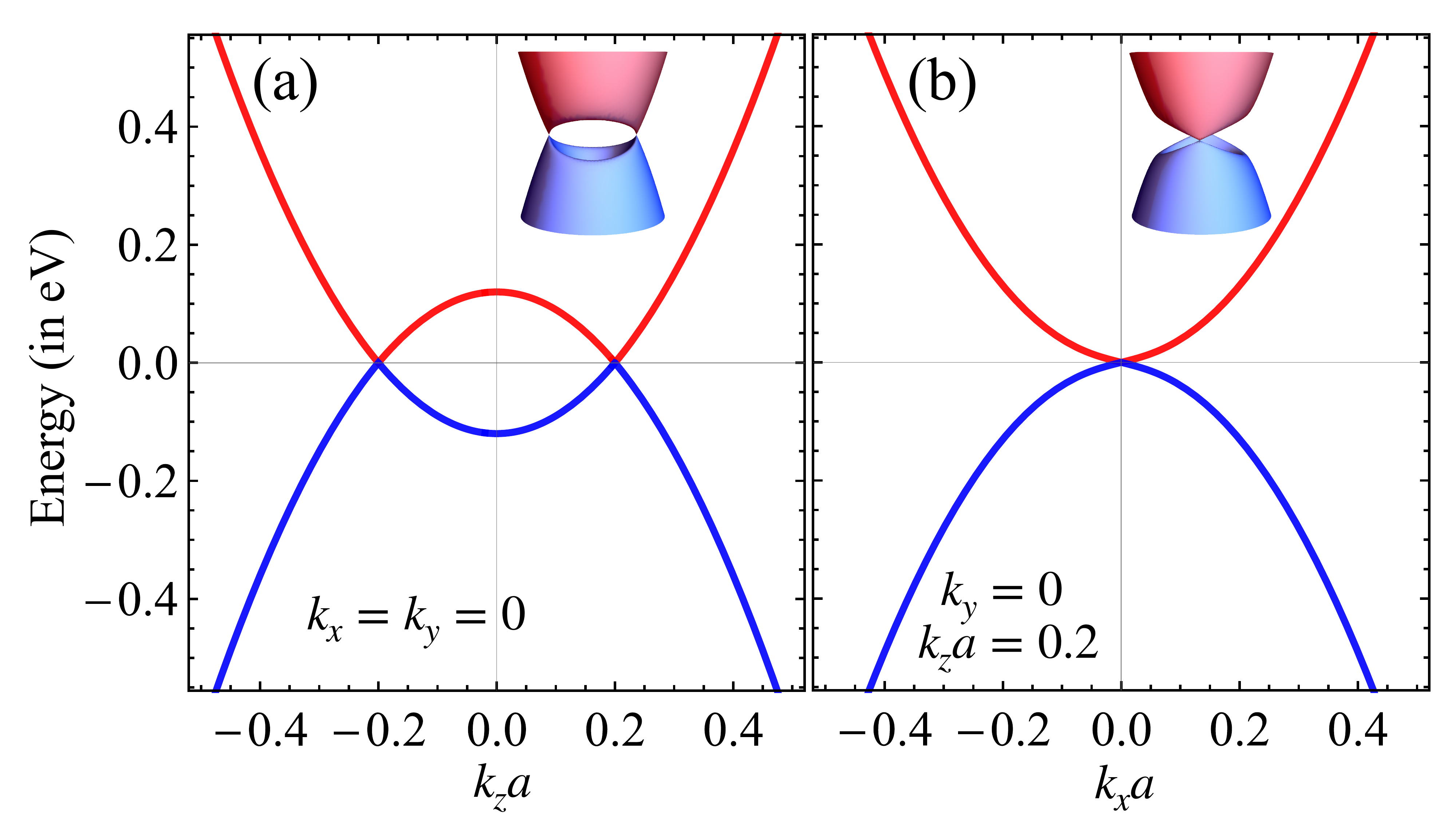}
\caption{Energy dispersion for Eq.~(\ref{ham}) as a function of (a) $k_z$ for $k_x=k_y = 0$, and (b) $k_x$ for $k_y = 0$ and $k_z = 0.2/a$. Here we have chosen $b = 0.12$\,eV, $t_1 = 0.25$\,eV,  $t_2 = 3$\,eV, and $a = 1\mathrm{\mathring{A}}$, which corresponds to the separation between Weyl nodes to be 0.4\,${\mathrm{\mathring{A}}}^{-1}$.}
\label{fig1}
\end{figure}
The corresponding eigenvectors are 
%
%
%
\begin{align}
\psi_{\lambda\bf k}=    \frac{1}{\sqrt{\mathcal{N}}}\left(\frac{e^{-i \phi_{\bf k} } \left(\chi +\lambda\sqrt{t_1^2a^2k_p^2+\chi ^2}\right)}{ak_p t_1 },1\right)^T~,
\end{align}
where we have denoted $\mathcal{N}=\frac{\left(\chi +\lambda\sqrt{t_1^2a^2k_p^2+\chi ^2} \right)^2}{t_1^2a^2k_p^2}+1$ and $\chi = t_2a^2(k_p^2+k_z^2)-b$ for convenience.

The energy dispersion depicting the two Weyl nodes is shown in Fig.~\ref{fig1}(a) and Fig.~\ref{fig1}(b). The parameters $t_2$ and $b$ define the band curvature and the distance between the Weyl points in the $k_z$ direction, whereas the parameter $t_1$ determines the slope in the vicinity of the Weyl points along the $k_p$ direction. The low-frequency optical response is dominated by the contribution from carrier excitations close to the chemical potential. This will become clear in the next section, where we compute optical conductivity in the linear response regime. 
\section{Optical conductivity}\label{Op_con}
In the weak-field limit or the linear response regime, the optical conductivity can be computed using the Kubo framework, where the frequency-dependent elements of the conductivity tensor are given by 
\begin{eqnarray}\label{kubo}
\sigma_{\alpha\beta}\left(\omega\right) &=& \frac{gi\hbar}{V} \sum_{mn}\frac{f_n-f_m}{\varepsilon_m-\varepsilon_n}\frac{\langle \psi_n | \hat j_{\alpha}| \psi_m\rangle \langle \psi_m | \hat j_{\beta}| \psi_n\rangle}{\hbar\omega + i\Gamma + \varepsilon_n-\varepsilon_m},
\end{eqnarray}
where $m$ and $n$ are quantum state indices which contain information about the band, spin, and the three-dimensional momentum ${\bf k}$. Furthermore, $f_{n/m} = \left(1+e^{\beta (\varepsilon_{n/m}-\mu)}\right)^{-1}$ is the Fermi-Dirac distribution function with $\beta = 1/k_\mathrm{B}T$ where $T$ is temperature and $\mu$ is the chemical potential.
The current operator is calculated as $\hat j_{\alpha/\beta} = e\hbar^{-1}\partial_{k_{\alpha/\beta}}\hat {\mathcal H}$, with $\lbrace\alpha,\beta\rbrace = \lbrace x, y, z\rbrace$. The spin degeneracy factor $g$ can be taken outside of the summation. To incorporate interaction and scattering effects at a phenomenological level, we have introduced a damping term, $\Gamma > 0$. It is important to note that scattering effects depend on the optical frequency and temperature, among other things. In particular, the intraband conductivity sector is largely affected by scattering on phonons and impurities. Therefore, in our modeling, we assume the intraband scattering term to be $\gamma_0 + \gamma T$, where $\gamma_0$ refers to the impurity contribution and the second term describes the temperature dependence due to phonons.\cite{TaAs_phonon, Kenneth1, Narang, Kenneth2} On the other hand, the interband scattering rate is assumed to be constant for simplicity. 
\subsection{Longitudinal conductivity: $\sigma_{xx}, \sigma_{yy}$}
\begin{figure}[ht!]
\includegraphics[width = \linewidth]{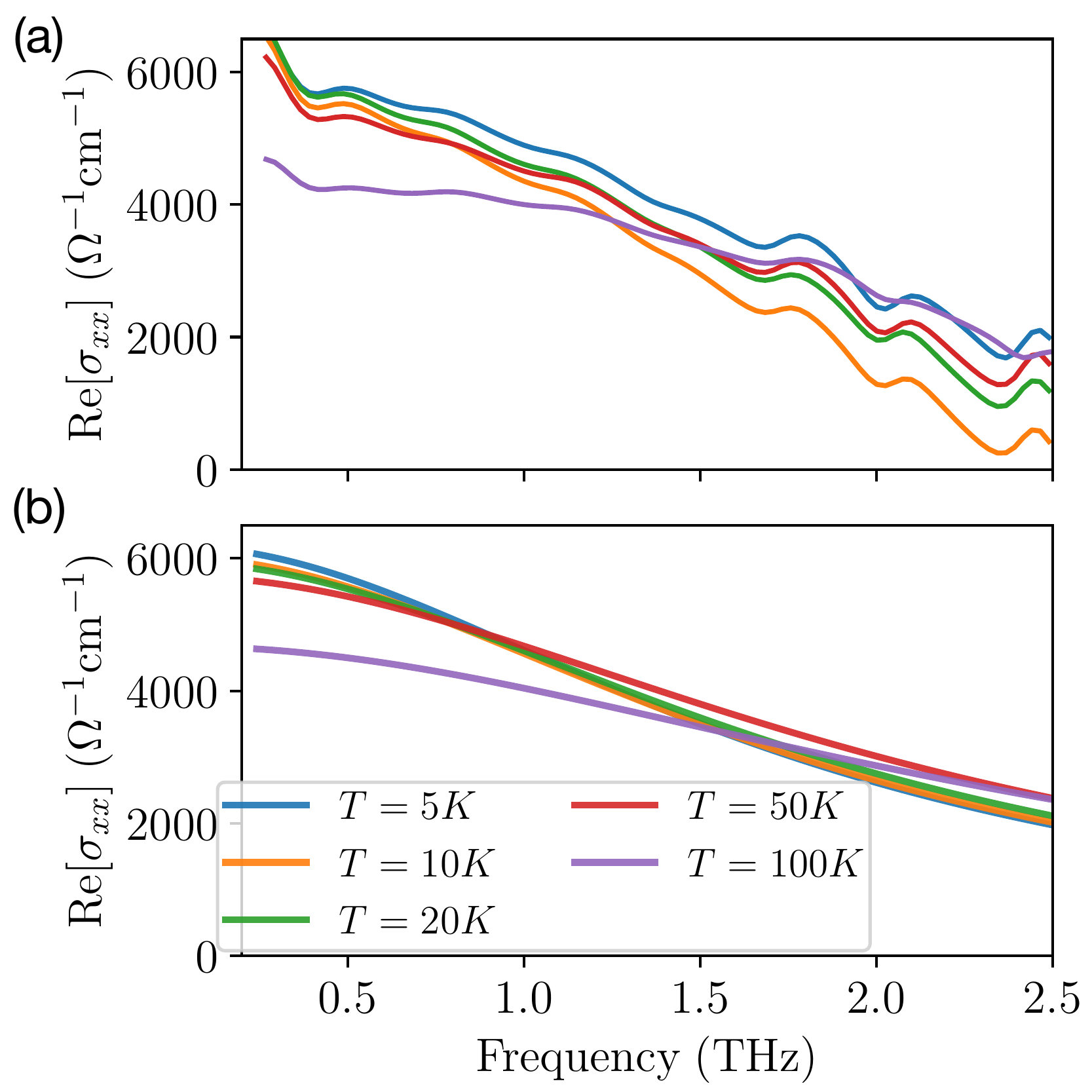}
\caption{Real part of the longitudinal optical conductivity as a function of frequency for five different temperatures as (a) measured experimentally and (b) computed using the model Hamiltonian given in Eq.~\eqref{ham} with parameters $a = 1\,\mathrm{\mathring{A}}$, $b = 0.12$\,eV, $t_1 = 0.25$\,eV,  $t_2 = 3$\,eV, $\mu = 43$\,meV, $\gamma_0 = 7$\,meV, and $\gamma = 0.034$\,meV/K.}
\label{fig2}
\end{figure}
From the Hamiltonian in Eq.~\eqref{ham}, we can obtain the current operator for the $\alpha$th ($\alpha\in(x,y)$) component at the given $\bf k$ as
\begin{align}
\hat j_{\alpha} = \frac{ea}{\hbar}\left(t_1\sigma_{\alpha} + 2t_2ak_{\alpha}\sigma_z\right)~.
\end{align}
Note again that $at_1/\hbar$ and $at_2/\hbar$ have the dimension of Fermi velocity \cite{RevModPhys.81.109}.  Since for the parameters relevant to the experiment the frequencies $\hbar \omega \ll \mu$, the contribution of interband transitions to the optical conductivity turned out to be negligibly small, and we neglect it for simplicity in this subsection. Furthermore, the azimuthal symmetry (see Eq.~\eqref{J_x_phi}) inherent in the Hamiltonian yields $\sigma_{xx}=\sigma_{yy}$. 

In Fig.~\ref{fig2}(a), we plot the real part of the optical conductivity obtained from THz time-domain spectroscopy (THz-TDS) measurements; see Appendix for details. Here, the oscillations are an artifact of the fast Fourier transform (FFT) after applying a step-function cutoff to remove back-reflection in the time-domain THz signal. In Fig.~\ref{fig2}(b), we present our simulation results, which show good qualitative agreement with the experiment. We consider a single pair of Weyl nodes to compute the optical conductivity and multiply the result by a factor of 3. The chemical potential is kept at $43$\,meV, which is close to the values reported in recent studies.\cite{Liu2018,Okamura2020} In the $\omega\to 0$ limit  the optical conductivity scales $\propto \mu^2/(\gamma_0+\gamma T)$. It is important to point out that the chemical potential also depends on the temperature,\cite{PhysRevB.93.085426, PhysRevB.103.075114} as can be seen in Fig.~\ref{mu_T} below. However, the leading-order temperature effect that qualitatively captures the data is due to the temperature dependence in the scattering rate, which originates from the temperature dependence of the interaction between electrons and various phononic modes.\cite{TaAs_phonon, Kenneth1, Narang, Kenneth2} 
\subsection{Transverse conductivity: $\sigma_{xy}$}
One of the reasons for the recent immense interest in Co$_3$Sn$_2$S$_2$ is its large AHE. According to previous studies,\cite{Liu2018, Chen} such a robust effect is guaranteed by the presence of three pairs of Weyl nodes, with each positive and negative chirality node separated roughly by $0.3\, \mathrm{\mathring{A}}^{-1}$ in momentum space. Here, we compute the transverse optical conductivity using a pair of Weyl nodes from the minimal model in Eq.~\eqref{ham}, and we again multiply the result by a factor of three to account for the three pairs. The transverse (Hall) conductivity is determined only by interband transitions, as shown in the Appendix. We find that the best fit to the experimental plots in Fig.~\ref{fig2} is obtained when the nodal separation in our effective two-band model is $\sim 0.4\, \mathrm{\mathring{A}}^{-1}$.

\begin{figure}[ht!]
\includegraphics[width = \linewidth]{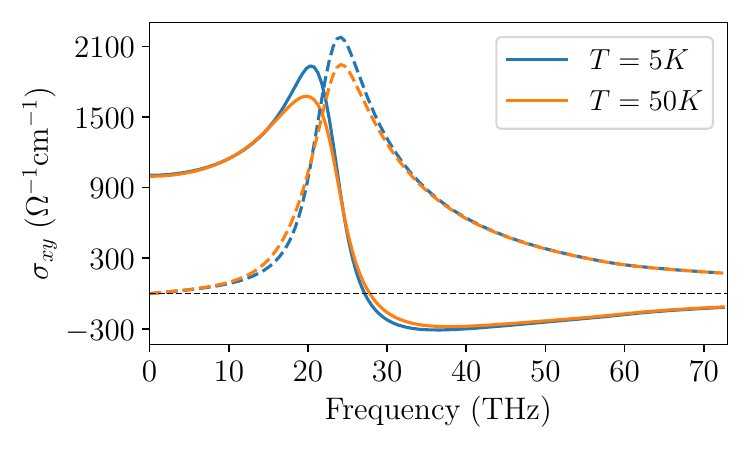}
\caption{Real (solid) and imaginary (dashed) parts of the transverse optical conductivity as a function of frequency for two different temperatures. Here, the interband scattering energy is  $\Gamma_{\rm inter} = 10$\,meV, whereas the rest of the parameters are the same as in Fig.~\ref{fig2}.}
\label{fig3}
\end{figure}

In Fig.~\ref{fig3} we plot the real and imaginary parts of the transverse optical conductivity as a function of frequency. In the limit $\omega\to 0$, the real part saturates to $\approx 1000~\Omega^{-1}\rm cm^{-1}$, which is close to the value reported in the literature.\cite{} It is important to note that this saturation value depends on the chemical potential, impurity density, interaction effects, and temperature. With increasing frequency, the real part increases and reaches a peak when the photon energy approaches twice the chemical potential, which is $2 \mu = 86$ meV. The conductivity decreases with further increase in frequency and remains negative at higher frequencies. On the other hand, the imaginary part of the conductivity vanishes in the $\omega\to 0$ limit. It sharply increases as the photon energy exceeds $86$ meV, then asymptotically approaches zero without changing sign. For smaller values of temperature and $\Gamma_{\rm in}$, the peaks become sharper and narrower. We emphasize that all these features shown in Fig.~\ref{fig3} agree well with recent experiments.\cite{Okamura2020}

%
\section{Faraday rotation}\label{Farday_angle}
Faraday rotation angle for a film of thickness $d$ as a function of frequency is given by~\cite{Okamura2020}  
\begin{align}\label{Fday}
    \Theta_\mathrm{F}(\omega) = \frac{Z_0d\sigma_{xy}(\omega)}{Z_0d\sigma_{xx}(\omega) + 1+n_\mathrm{s}}~,
\end{align}
where $n_\mathrm{s}$ is the refractive index of the substrate and $Z_0 \approx 377~\Omega$ is the vacuum impedance.

\begin{figure}[ht!]
\includegraphics[width = \linewidth]{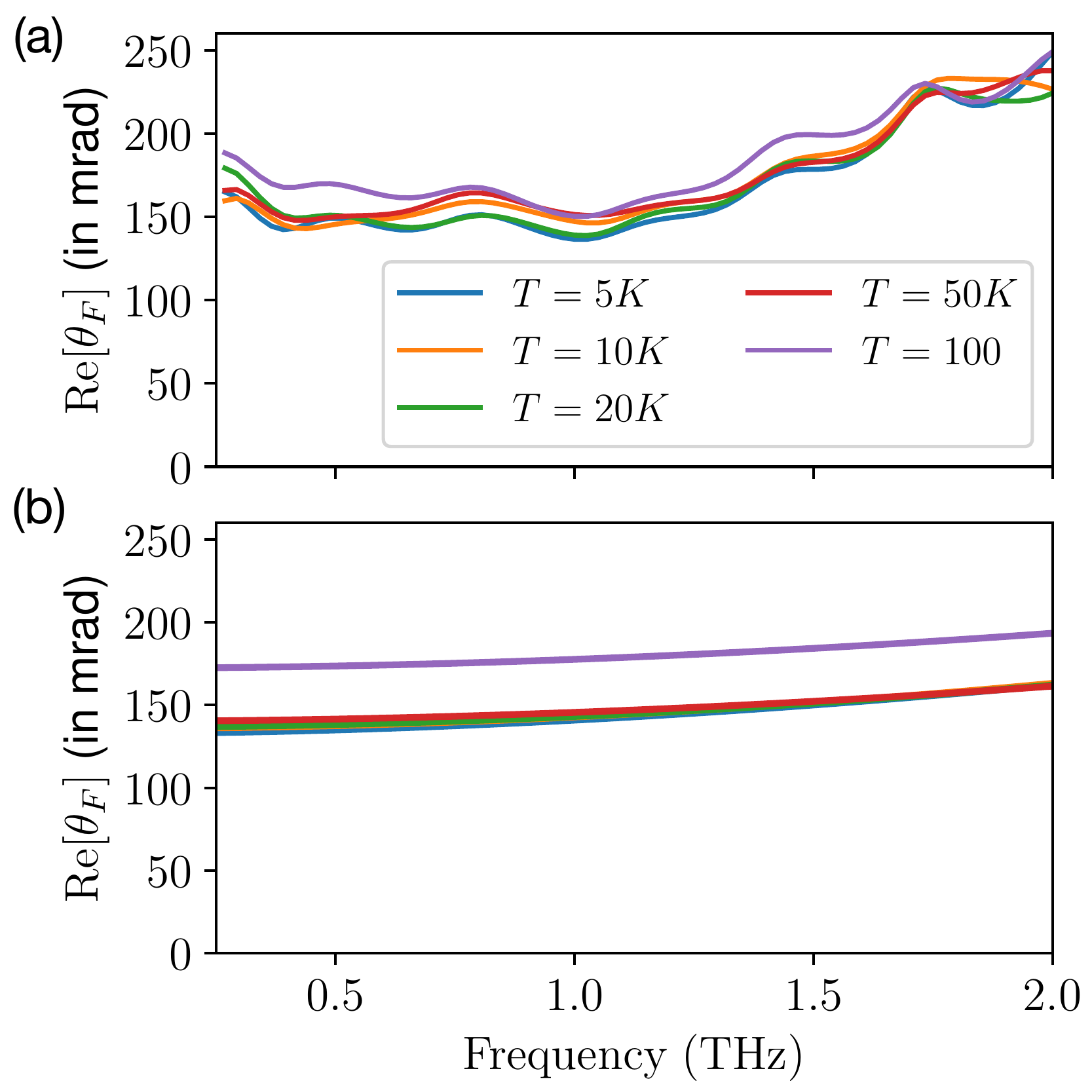}
\caption{Real part of the Faraday angle as a function of frequency for different values of temperature from (a)~experimental data and (b)~Eq.~\eqref{ham}. The film thickness is $d=50$\,nm, and the substrate refractive index is $n_\mathrm{s} = 1.6$. Other parameters are the same as in Fig.~\ref{fig2}.}
\label{fig4}
\end{figure}

In Fig.~\ref{fig4}, we plot the experimental data for the real part of the Faraday angle as a function of frequency for five values of temperature and compare them with theoretical calculations. Here again, the oscillations seen in measured spectra are a FFT artifact. The real part of the Faraday angle can also be written as \\
\\
${\rm Re}[\Theta_\mathrm{F}(\omega)]$
\begin{align}
   = \frac{{\rm Im}[\sigma_{xx}(\omega)] {\rm Im}[\sigma_{xy}(\omega)]+{\rm Re}[\tilde\sigma_{xx}(\omega)] {\rm Re}[\sigma_{xy}(\omega)]}{{\rm Im}[\sigma_{xx}(\omega)]^2+{\rm Re}[\tilde\sigma_{xx}(\omega)]^2}~,
\end{align}
where ${\rm Re}[\tilde\sigma_{xx}(\omega)] = {\rm Re}[\sigma_{xx}(\omega)] + (1+n_\mathrm{s})/(Z_0d)$. The term $(1+n_\mathrm{s})/(Z_0d)$ is independent of frequency and provides only an overall shift to the real part of the longitudinal conductivity. 
For the given set of parameters, we can see from Fig.~\ref{fig3} that ${\rm Re}[\sigma_{xy}(\omega)]\gg{\rm Im}[\sigma_{xy}(\omega)]$ in the low frequency regime. Furthermore, ${\rm Re}[\tilde\sigma_{xx}(\omega)]$ remains larger than ${\rm Im}[\sigma_{xx}(\omega)]$ throughout the frequency range (not shown). Therefore, for qualitative understanding, we can simplify Eq.~\eqref{Fday} such that ${\rm Re}[\Theta_\mathrm{F}(\omega)]\approx {\rm Re}[\sigma_{xy}(\omega)]/{\rm Re}[\tilde\sigma_{xx}(\omega)]$. In the low frequency range,  ${\rm Re}[\sigma_{xy}(\omega)]$ remains roughly constant, since the main contributing factor here is the Weyl node separation. 
Therefore, ${\rm Re}[\sigma_{xy}(\omega)]/{\rm Re}[\tilde\sigma_{xx}(\omega)]$ increases with increasing frequency because ${\rm Re}[\tilde\sigma_{xx}(\omega)]$ decreases. As the temperature is  increased, ${\rm Re}[\tilde\sigma_{xx}(\omega)]$ decreases, which results in larger ${\rm Re}[\Theta_\mathrm{F}(\omega)]$ as shown in Fig.~\ref{fig4}. This non-uniform temperature dependence is attributed to the interplay between the temperature dependence of the doping profile and the scattering rates. 

\begin{figure}[t!]
\includegraphics[width = 0.9\linewidth]{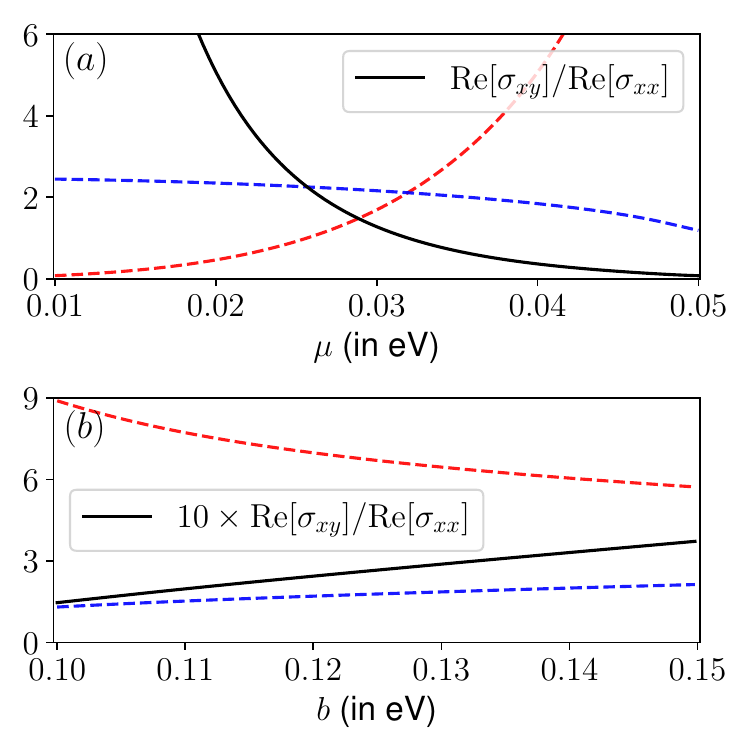}
\caption{Ratio of real parts of the transverse and longitudinal optical conductivity as a function of (a)~chemical potential, $\mu$, and (b)~energy parameter $b$, which dictates the Weyl node separation for a given $a$ and $t_2$. In part (b), the ratio has been multiplied by a factor of 10. Dashed blue and red lines represent ${\rm Re}[\sigma_{xy}]$ and ${\rm Re}[\sigma_{xx}]$ respectively in the units of $3e^2/(2\pi^2 h a).$ We have fixed $T\sim 12$ K for these plots and the rest of the parameters are the same as in Fig.~\ref{fig2}.}
\label{fig6}
\end{figure}

Figure \ref{fig6} illustrates the constraints imposed by the model on the values of the chemical potential $\mu$ and the energy parameter $b$. While the real part of transverse conductivity changes slowly with increase in $\mu$, the longitudinal conductivity increases sharply ($\propto \mu^2$), which results in a sharp decline of the ratio ${\rm Re}[\sigma_{xy}]/{\rm Re}[\sigma_{xx}]$. Similarly, from Fig.~\ref{fig6}(b) it is obvious that ${\rm Re}[\sigma_{xx}]$ strongly depends on $b$. Both features are needed to explain the experimental data.

\section{Conclusions}\label{Conclusion}
Time-reversal-symmetry-broken Weyl semimetals promise a plethora of interesting gyrotropic effects in the low-frequency optical response, which can be traced to Weyl node separation and are therefore of topological origin. Moreover, the presence of topologically protected surface states makes these systems promising candidates for quantum information applications. Because of these interesting properties, the search for materials, mainly driven by {\it ab initio} simulations followed by experimental discoveries, suggested several magnetic Weyl semimetal candidates.

Co$_3$Sn$_2$S$_2$ reportedly has one of the largest DC AHE ($>1000~\Omega^{-1}{\rm cm}^{-1}$) among known Weyl semimetals. In this work, we have carried out a systematic experimental and theoretical study of the optical response in Co$_3$Sn$_2$S$_2$ in the THz frequency range. The THz spectroscopy is able to probe the most interesting region of low-energy electron excitations in the vicinity of Weyl nodes. It is also particularly suitable for analyzing with an effective low-energy model, which provides an intuitive physical interpretation of the experimental data based on a limited number of material parameters. Using this model, we computed the longitudinal and transverse optical conductivities and the Faraday rotation angle within the Kubo formalism. Given the large temperature variation, we also accounted for the temperature-dependent chemical potential. We demonstrated that our low-energy model presents a consistent physical picture and provides a good qualitative explanation of the experimental data including the real parts of the longitudinal conductivity and the Faraday angle, as well as  the temperature variation. 


The combination of THz spectroscopy and its interpretation in terms of the low-energy model imposes stringent constraints on the values of material parameters. This approach complements well the existing and future first-principle studies and can be extended to other magnetic Weyl semimetals.

\section{ACKNOWLEDGEMENTS}
J.K.\ acknowledges support from the U.S.\ Army Research Office (through Award No.\ W911NF-21-1-0157, W911NF-23-1-0410, W911NF-25-2-0150), the Gordon and Betty Moore Foundation (through Grant No.\ 11520), and the Robert A.\ Welch Foundation (through Grant No.\ C-1509). 

\appendix
\section{Conductivity calculation}
Using $k_x\to k_p\cos\phi_{\bf k}$ and $k_y\to k_p\sin\phi_{\bf k}$, Eq.~\eqref{ham} can be written as
\begin{align}
\hat{\mathcal{H}} = \left(
\begin{array}{cc}
 a^2 t_2 \left(k_p^2+k_z^2\right)-b & a k_p t_1 e^{-i \phi_{\bf k} } \\
 a k_p t_1 e^{i \phi_{\bf k} } & b-a^2 t_2 \left(k_p^2+k_z^2\right) \\
\end{array}
\right)~.
\end{align}
The eigenstates are 
\begin{align}\nonumber
\psi_{v\bf k}=\begin{pmatrix}\frac{e^{-i \phi_{\bf k} } \left(a^2 t_2 \left(k_p^2+k_z^2\right)-b-\frac{\hbar\omega_{\bf k}}{2}\right)}{a k_p t_1 \sqrt{\frac{\left(a^2 t_2 \left(-\left(k_p^2+k_z^2\right)\right)+b+\frac{\hbar\omega_{\bf k}}{2}\right)^2}{a^2 k_p^2 t_1^2}+1}}\\\frac{1}{\sqrt{\frac{\left(a^2 t_2 \left(-\left(k_p^2+k_z^2\right)\right)+b+\frac{\hbar\omega_{\bf k}}{2}\right)^2}{a^2 k_p^2 t_1^2}+1}}\end{pmatrix}~,\\
\end{align}
\begin{align}
\psi_{c\bf k}=\begin{pmatrix}\frac{e^{-i \phi_{\bf k} } \left(a^2 t_2 \left(k_p^2+k_z^2\right)-b+\frac{\hbar\omega_{\bf k}}{2}\right)}{a k_p t_1 \sqrt{\frac{\left(a^2 t_2 \left(-\left(k_p^2+k_z^2\right)\right)+b-\frac{\hbar\omega_{\bf k}}{2}\right)^2}{a^2 k_p^2 t_1^2}+1}}\\\frac{1}{\sqrt{\frac{\left(a^2 t_2 \left(-\left(k_p^2+k_z^2\right)\right)+b-\frac{\hbar\omega_{\bf k}}{2}\right)^2}{a^2 k_p^2 t_1^2}+1}}\end{pmatrix}~,
\end{align}
where $\hbar\omega_{\bf k} =2 \sqrt{\left(b-a^2 t_2 \left(k_p^2+k_z^2\right)\right)^2+a^2 k_p^2 t_1^2}$.
Current operators in the cylindrical polar coordinate are given as  
\begin{align}
j_{x\bf k} = \frac{e}{\hbar}\left(
\begin{array}{cc}
 2 a^2 k_p t_2 \cos\phi_{\bf k} & a t_1 \\
 a t_1 & -2 a^2 k_p t_2 \cos\phi_{\bf k} \\
\end{array}
\right)~,
\end{align}
\begin{align}
j_{y\bf k} = \frac{e}{\hbar}\left(
\begin{array}{cc}
 2 a^2 k_p t_2 \sin\phi_{\bf k} & -i a t_1 \\
 i a t_1 & -2 a^2 k_p t_2 \sin\phi_{\bf k} \\
\end{array}
\right)~,
\end{align}
and
\begin{align}
j_{z\bf k} = \frac{e}{\hbar}\left(
\begin{array}{cc}
 2 a^2 k_z t_2 & 0 \\
 0 & -2 a^2 k_z t_2 \\
\end{array}
\right)~.
\end{align}
The current matrix elements are 
\begin{align}
\langle \psi_{c\bf k}| j_{x\bf k} | \psi_{c\bf k}\rangle = \frac{e}{\hbar}\frac{a^2 k_p \cos\phi_{\bf k} \left(2 t_2 \left(a^2 t_2 \left(k_p^2+k_z^2\right)-b\right)+t_1^2\right)}{\sqrt{\left(b-a^2 t_2 \left(k_p^2+k_z^2\right)\right)^2+a^2 k_p^2 t_1^2}}~,\end{align}
\begin{align}
\langle \psi_{v{\bf k}}| j_{x{\bf k}} | \psi_{v{\bf k}}\rangle = -\langle \psi_{c{\bf k}}| j_{x\bf k} | \psi_{c\bf k}\rangle~,
\end{align}
\\
$\langle \psi_{v\bf k}| j_{x\bf k} | \psi_{c\bf k}\rangle =$
\begin{align}
-\frac{e}{\hbar}\frac{-2 \cos\phi_{\bf k} \left(a^2 t_2 \left(k_p^2-k_z^2\right)+b\right)-i \hbar\omega_{\bf k} \sin\phi_{\bf k}}{2 k_p \sqrt{\frac{\hbar\omega_{\bf k}}{-2 a^2 t_2 \left(k_p^2+k_z^2\right)+2 b+\hbar\omega_{\bf k}}} \sqrt{\frac{\hbar\omega_{\bf k}}{2 a^2 t_2 \left(k_p^2+k_z^2\right)-2 b+\hbar\omega_{\bf k}}}}~,
\end{align}
and 
\begin{align}
\langle \psi_{c\bf k}| j_{x\bf k} | \psi_{v\bf k}\rangle =(\langle \psi_{v\bf k}| j_{x\bf k} | \psi_{c\bf k}\rangle)^{*}~.
\end{align}
Since the energies do not depend on $\phi_{\bf k}$, polar integration only acts on the square of the dipole elements. We have\\
\\
$\int_{0}^{2\pi} ~d\phi_{\bf k} |\langle \psi_{c\bf k}| j_{x\bf k} | \psi_{c\bf k}\rangle|^2$
\begin{align}\label{J_x_phi}
 = \frac{e^2}{\hbar^2}\frac{\pi  a^4 k_p^2 \left(2 t_2 \left(a^2 t_2 \left(k_p^2+k_z^2\right)-b\right)+t_1^2\right)^2}{\left(b-a^2 t_2 \left(k_p^2+k_z^2\right)\right)^2+a^2 k_p^2 t_1^2}~,
 \end{align}
and\\
\\
$\int_{0}^{2\pi} ~d\phi_{\bf k} |\langle \psi_{c\bf k}|j_{x\bf k} | \psi_{v\bf k}\rangle|^2$
\begin{align}
  = \frac{e^2}{\hbar^2}\frac{\pi  a^2 t_1^2 \left(-4 a^2 b k_z^2 t_2+2 a^4 t_2^2 \left(k_p^4+k_z^4\right)+a^2 k_p^2 t_1^2+2 b^2\right)}{\left(b-a^2 t_2 \left(k_p^2+k_z^2\right)\right)^2+a^2 k_p^2 t_1^2}~.
\end{align}
From Eq.~\eqref{kubo}, we have\\
\\
$\dfrac{\sigma_{xx}\left(\omega\right)}{gi\hbar/(8\pi^3)}$
\begin{eqnarray}\nonumber
 = \int k_pdk_pdk_z \bigg(\frac{f_{v\bf k}-f_{c\bf k}}{\hbar\omega_{\bf k}}\int_{0}^{2\pi} ~d\phi_{\bf k} |\langle \psi_{c\bf k}| j_{x\bf k} | \psi_{v\bf k}\rangle|^2\\\nonumber
\times\left[\frac{1}{\hbar\omega + i\Gamma_{\rm inter} + \hbar\omega_{\bf k}}
+\frac{1}{\hbar\omega + i\Gamma_{\rm inter} - \hbar\omega_{\bf k}}\right]\\\nonumber
+\left(-\frac{\partial f_{v\bf k}}{\partial\varepsilon_{v\bf k}} - \frac{\partial f_{c\bf k}}{\partial\varepsilon_{c\bf k}}\right)\int_{0}^{2\pi} d\phi_{\bf k} ~\frac{|\langle \psi_{c\bf k}| j_{x\bf k} | \psi_{c\bf k}\rangle|^2}{\hbar\omega + i\Gamma_{\rm intra}}\bigg).\\
\end{eqnarray}
Plugging the expressions for computed polar integrals and by taking appropriate set of parameters, the rest of the integration we perform numerically. 
%
%
%
%
\subsection{Transverse conductivity: $\sigma_{xy}$}
Next we look at the Hall term, which only gets interband contribution. This is due to the fact that 
\begin{align}
  \langle \psi_{c\bf k}| j_{x\bf k} | \psi_{c\bf k}\rangle\langle \psi_{c\bf k}| j_{y\bf k} | \psi_{c\bf k}\rangle \propto \sin\phi_{\bf k}\cos\phi_{\bf k}~,
\end{align}
such that the polar integration vanishes. Thus we can write 
\begin{eqnarray}\nonumber
\frac{\sigma_{xy}\left(\omega\right)}{gi\hbar/(8\pi^3)} = \int d^3k\frac{f_{v\bf k}-f_{c\bf k}}{\hbar\omega_{\bf k}}\frac{\langle \psi_{c\bf k} | j_{x\bf k}| \psi_{v\bf k}\rangle\langle \psi_{v\bf k} | j_{y\bf k}| \psi_{c\bf k}\rangle }{\hbar\omega + i\Gamma_{\rm inter} + \hbar\omega_{\bf k}}\\\nonumber
+\frac{f_{v\bf k}-f_{c\bf k}}{\hbar\omega_{\bf k}}\frac{\langle \psi_{v\bf k} | j_{x\bf k}| \psi_{c\bf k}\rangle \langle \psi_{c\bf k} | j_{y\bf k}| \psi_{v\bf k}\rangle }{\hbar\omega + i\Gamma_{\rm inter} - \hbar\omega_{\bf k}}~.\\
\end{eqnarray}
The dipole matrix element relevant in this case is 
$\langle \psi_{v\bf k}| j_{y\bf k} | \psi_{c\bf k}\rangle$
\begin{align}
 =\frac{e}{\hbar}\frac{-2 \sin\phi_{\bf k} \left(a^2 t_2 \left(k_p^2-k_z^2\right)+b\right)+i \hbar\omega_{\bf k} \cos\phi_{\bf k}}{2 k_p \sqrt{\frac{\hbar\omega_{\bf k}}{-2 a^2 t_2 \left(k_p^2+k_z^2\right)+2 b+\hbar\omega_{\bf k}}} \sqrt{\frac{\hbar\omega_{\bf k}}{2 a^2 t_2 \left(k_p^2+k_z^2\right)-2 b+\hbar\omega_{\bf k}}}}~.
\end{align}
and 
\begin{align}
\langle \psi_{c\bf k}| j_{y\bf k} | \psi_{v\bf k}\rangle =(\langle \psi_{v\bf k}| j_{y\bf k} | \psi_{c\bf k}\rangle)^{*}~.
\end{align}
Using these dipole matrix elements, the polar angle integral can be computed and we get
\begin{align}\nonumber
\int_{0}^{2\pi} ~d\phi_{\bf k} \langle \psi_{c\bf k}| j_{x\bf k} | \psi_{v\bf k}\rangle\langle \psi_{v\bf k}| j_{y\bf k}| \psi_{c\bf k}\rangle\\
 = \frac{2 i a^2 t_1^2 \left(a^2 t_2 (k_p-k_z) (k_p+k_z)+b\right)}{\sqrt{\left(b-a^2 t_2 \left(k_p^2+k_z^2\right)\right)^2+a^2 k_p^2 t_1^2}}~.
 \end{align}
\subsection{Temperature dependence of chemical potential}
Since energy dispersion is linear close to the Weyl nodes, the density of states $N(\varepsilon)$ varies as $\varepsilon^2$. In this case, we expect the temperature effects to be important. Firstly, the chemical potential becomes a function of temperature. This is because the total number of charge carriers is conserved as we change the temperature. 
The expression for the temperature dependent chemical potential for a linearly dispersing Weyl node is given as 
\begin{align}\label{mu_TA}
\mu(\tilde T) = \frac{\sqrt[3]{2} \left(9 \mu_0^3+\sqrt{81 \mu_0^6+12 \pi ^6 \tilde T^6}\right)^{2/3}-2 \sqrt[3]{3} \pi ^2 \tilde T^2}{6^{2/3} \sqrt[3]{9 \mu_0^3+\sqrt{81 \mu_0^6+12 \pi ^6 \tilde T^6}}}
\end{align}
where $\tilde T = k_\mathrm{B}T$, and $\mu_0 = \mu(0)$ is the bare chemical potential. In the limiting cases we have simpler expressions,
\begin{figure}[t!]
\includegraphics[width = \linewidth]{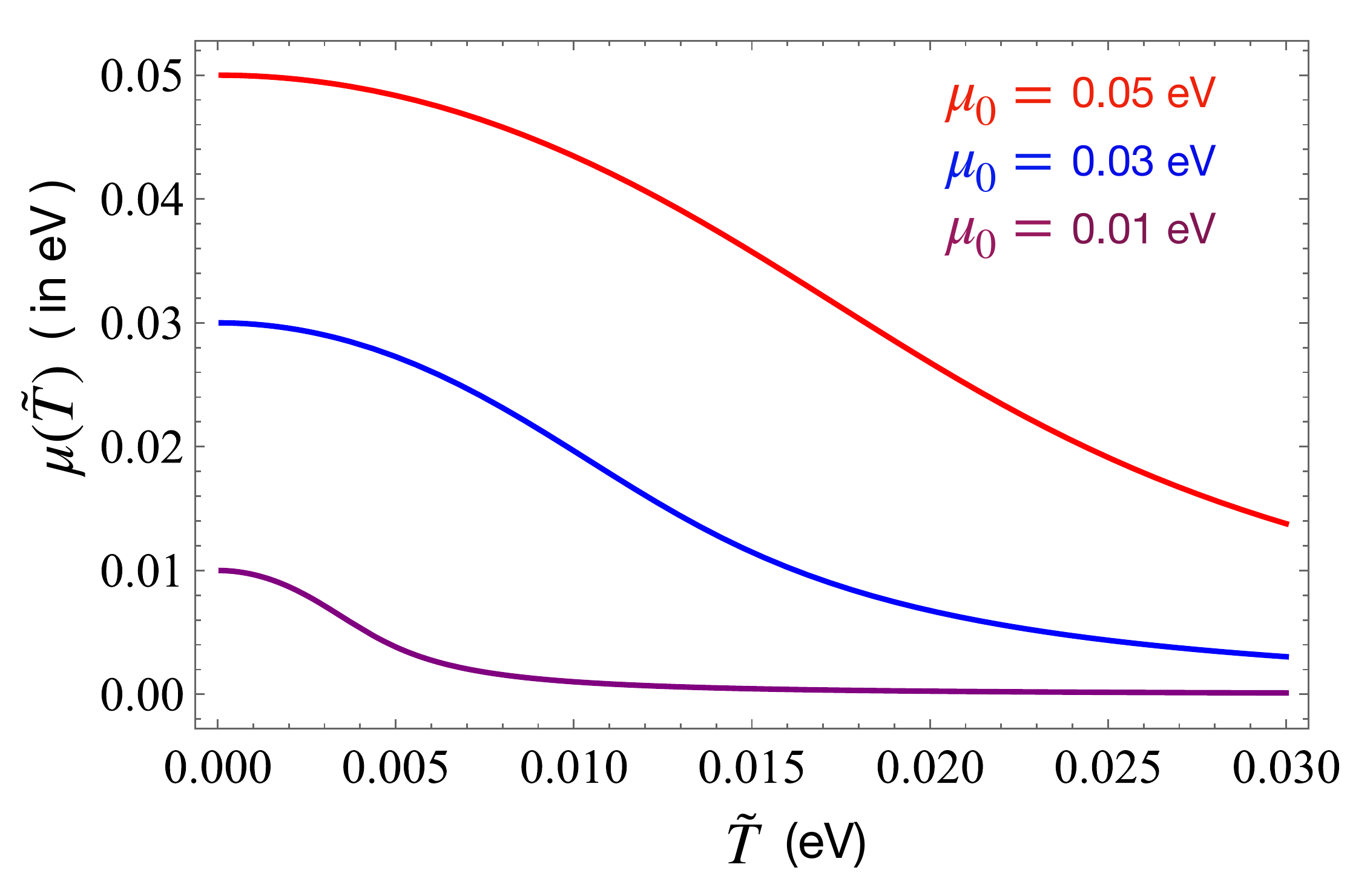}
\caption{Temperature dependence of chemical potential for three values of $\mu_0$.}
\label{mu_T}
\end{figure}
\begin{align}
\mu(\tilde T) = \mu_0 - \frac{\pi^2}{3}\frac{\tilde T^2}{\mu_0},~ \mu_0 \gg \tilde T, \\
\mu(\tilde T) = \frac{\mu_0^3}{\pi^2\tilde T^2},~ \mu_0 \ll \tilde T. 
\end{align}
Although these results are obtained for one node and in the linear band limit, for chemical potential not far away from the nodes, we can still utilize Eq.~\eqref{mu_TA}. This facilitates our goal to capture the qualitative and order of magnitude quantitative results.

\section{Terahertz time-domain magnetospectroscopy}
The terahertz time-domain setup is described in detail in Ref.~\onlinecite{kim_multimode_2025}. It was a commercial TERAFLASH PRO, a time-domain terahertz spectroscopy platform from TOPTICA Photonics. The low-temperature magnetospectroscopy measurements were conducted in a Lakeshore DryMag cryogen-free system. During the sample cooldown, we applied a 2\,T magnetic field at $T=250\,$K and switched it off at $T=50\,$K. All subsequent low-temperature measurements were conducted without any applied magnetic field. We utilized a THz linear polarizer oriented at \(+45^\circ\) and \(-45^\circ\) with respect to the input THz polarization to obtain the corresponding electric field $E_{+45^\circ}(\omega)$ and $E_{-45^\circ}(\omega)$. Then, the $x-$ and the $y-$component of the transmitted electric field could be calculated by $E_x(\omega)=E_{+45^\circ}(\omega) + E_{-45^\circ}(\omega)$ and $E_y(\omega)=E_{+45^\circ}(\omega) - E_{-45^\circ}(\omega)$, respectively.\cite{ChoiEtAl2019NM,Rodriguez-BarriosEtAl2025} Subsequently, the complex Faraday rotation angle $\Theta_\mathrm{F}(\omega)$ was obtained by $\Theta_\mathrm{F}(\omega) = \mathrm{arctan}\left( E_y(\omega)/E_x(\omega) \right)$.\cite{Okamura2020} We calculated the longitudinal complex THz conductivity $\sigma_{xx}(\omega)$ with the Tinkham equation,\cite{glover_conductivity_1957} using the time-domain data of a bare sapphire substrate as a reference. The complex Hall conductivity $\sigma_{xy}(\omega)$ and the longitudinal complex THz conductivity $\sigma_{xx}(\omega)$ are related through the Faraday rotation angle $\Theta_\mathrm{F}$ as shown in Eq.~\ref{Fday}.
\bibliography{Ref2,Hongjing_Co3Sn2S2}

\end{document}